\documentclass[a4paper,10pt]{article}

\usepackage[english]{babel}

\usepackage{amssymb,amsmath}
\usepackage{graphicx}
\usepackage{subcaption}
\usepackage{gensymb}
\graphicspath{{figs/}}
\usepackage[colorinlistoftodos]{todonotes}
\usepackage{xcolor}
\usepackage{xfrac}
\usepackage[normalem]{ulem}
\usepackage{tikz}
\usepackage{siunitx}
\usepackage{authblk}
\usepackage{cite}
\usepackage{tabularx}

\usepackage{hyperref}
\hypersetup{
	colorlinks,
	citecolor=blue,
	linkcolor=blue,
	urlcolor=blue
}

\newcommand{\av}[1]{\left\langle {#1} \right\rangle}
\newcommand{\avt}[1]{\overline{#1}}

\renewcommand{\vec}[1]{\mathbf{#1}}
\newcommand{\volume}{{\ooalign{\hfil$V$\hfil\cr\kern0.08em--\hfil\cr}}}

\usepackage{setspace}
\usepackage[margin=2.5cm]{geometry}

\title{Bubble-induced convection and flow-instability in a soft reactor}

\author[1]{Ron Shnapp}

\author[2]{Markus Holzner}

\affil[1]{\small{Ben Gurion University of the Negev, Mechanical engineering department, POB 653, Beer Sheva 8410501, Israel}}

\affil[2]{\small{Institute of Hydraulic Engineering and River Research, University of Natural Resources and Life Sciences, 1180 Vienna, Austria
}}

\date{}

\begin{document}

\onehalfspacing

\maketitle

\begin{abstract}
	Buoyancy-driven bubbly flows play pivotal roles in various scenarios, such as the oxygenation and mixing in the upper ocean and the reaction kinetics in chemical and bio-reactors. This work focuses on the convective flow induced by the localized release of large air bubbles ($D_b=3.7$ mm, $\mathrm{Re}_b=950$) in a water tank, exploring the resulting flow and the transition from laminar to disturbed states as a function of the Rayleigh number in the range $3\times10^3 \div 2\times10^5$. At low $\mathrm{Ra}$ the flow is smooth and laminar with weak temporal oscillations, while a highly disturbed state appears above a critical value $\mathrm{Ra}_c$. A theoretical analysis is presented that links the mean flow circulation to the Rayleigh number. Through an experimental investigation, utilizing 3D-particle tracking velocimetry and flow visualization, we confirm the theory presented, and characterize the laminar to disturbed transition in the system. The study offers insights into the convective flow dynamics generated by bubbles, with implications for applications such as bio-reactor soft mixer design.
\end{abstract}


\section{Introduction}\label{sec:intro}

Buoyancy forces acting on bubbles due to Archimedes' principle, can drive liquid flows. Such buoyancy-driven bubbly flows occur in many natural and industrial settings, such as in the upper ocean~\cite{Thorpe1992} and in chemical~\cite{Harteveld2003, Oresta2009, Mezui2022} or bio-~\cite{Liu2010} reactor, where they induce mixing of dissolved materials. Buoyancy driven flows appear due to the fundamental requirement that the spatial mass density distribution is non-uniform. In single phase flows, such differences often appear due to temperature or solute concentration differences, such as in the well-known Rayleigh-B\'{e}nnard~\cite{Lohse2010} or the horizontal convection~\cite{Hughes2008}. In multiphase systems, on the other hand, flows are driven by the interfacial forces, such as in liquids with dispersed gas bubbles. The present work focuses on the convective flow generated by releasing air bubbles at the bottom-side of a liquid talk.

The introduction of air bubbles to a liquid tank can lead to a vast variety of different phenomena depending on the particular parameter regime, as reviewed in Ref.~\cite{Lohse2018}. Our work focuses on the water flow generated by air bubbles whose diameters are in the few millimeters range. Such bubbles ascend through the water in plumes and their bubble Reynolds numbers are much larger than 1. As these bubbles rise through the liquid, agitation and vorticity is generated in their wakes~\cite{Almeras2015, Mathai2020}. At high volume fractions, a regime of strongly disturbed flow exists that bears some similarity to turbulence in the sense that it is chaotic and possesses a continuous kinetic energy spectrum~\cite{Risso2018, Innocenti2021}. Notably, the kinetic energy spectrum in these flows is steeper, scaling as -3 power or the wave number, and this regime initiates at scales on the order of the bubble diameter.

When bubbles are added non-uniformly to a flow, the uneven density distribution results in large scale convective flows. In bubble columns, uneven injection leads to highly chaotic flow patters with spatio-temporal fluctuations in the volume fraction~\cite{Gong2009, Mezui2023}. When such bubbles are released from a fixed position they result in unstable bubble plumes that carry the liquid upward with them~\cite{Caballina2003, Simiano2006}. An important characteristic of these buoyancy-driven bubbly flows is that they are very often unstable. For example, horizontal layers of bubbles are known to exhibit an instability with characteristics very similar to the Rayleigh-B\'{e}nnard instability convection~\cite{Climent1999, Ruzicka2003, Nakamura2021}.

The present work focuses on the convective flow generated in a water tank due to the introduction of air bubbles at the bottom of one of the side walls of the tank (Fig~\ref{fig:setup}a). This configuration is observed to generate a convective circulation across the whole tank, with characteristics similar to the cavity flows explored in Refs.~\cite{Rockwell1978, Koseff1984}. Furthermore, while at sufficiently low air flow rates the flow is laminar and smooth, above a certain threshold a new regime is reached in which strong fluctuations ensue at a range of scales. The system is explored experimentally by 3D-particle tracking velocimetry~\cite{Shnapp2019, Shnapp2022} and flow visualization. The experimental system is described in Sec.~\ref{sec:methods}. In Sec.~\ref{sec:euilibrium_theory} a theory is presented for the dependence of the mean flow circulation on the Rayleigh number which is controlled through the bubble injection rate. In Sec.~\ref{sec:results} experimental results are used to confirm our theory and to characterize the transition in this flow. Lastly, a brief summary and discussion are presented in Sec.~\ref{sec:conclusinos}.

% ===================================================
%                    Methods
% ===================================================

\section{Methods}\label{sec:methods}

\subsection{Experimental system}

\begin{figure}[h!]
	\centering
	\includegraphics[width=\textwidth]{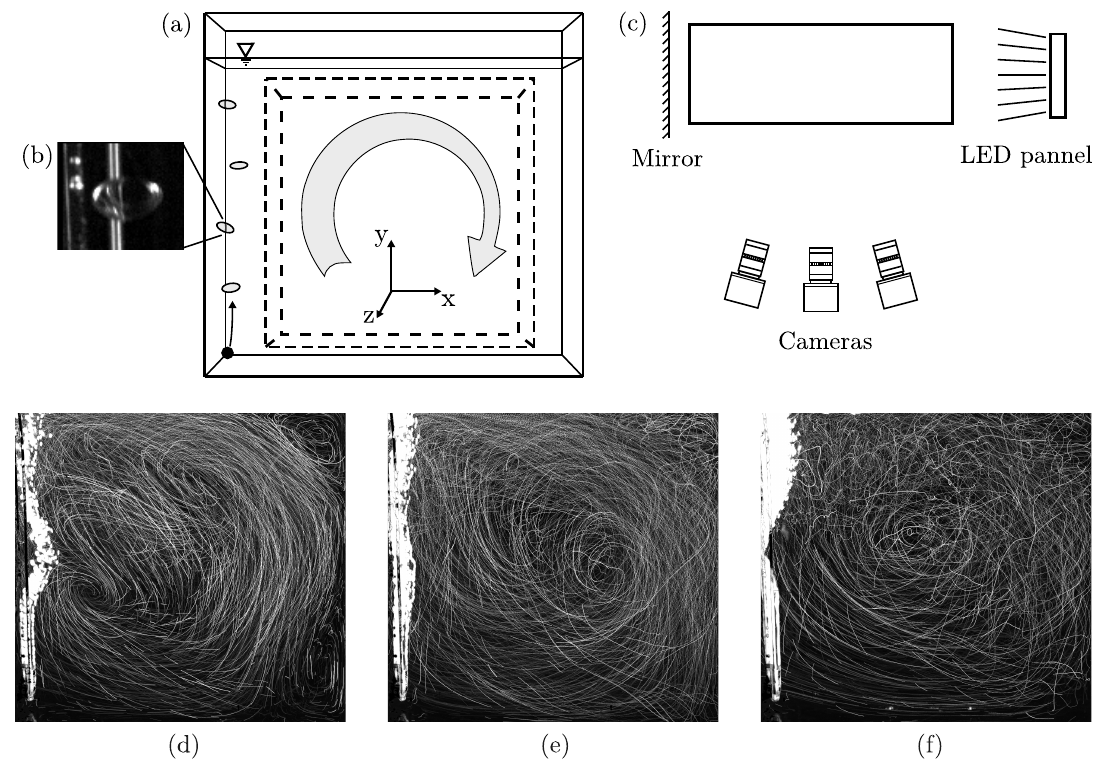}
	
	\caption{(a) Front view schematics of the experimental apparatus; the bubble source is shown as a black circle in the lower left corner, and the region of interest for the measurements is shown in dashed lines. (b) An image of one of a bubble from the experiment. (c) Top view schematic of the experiment measurement system. (d), (e) and (f) Flow visualization made by overlaying experimental images. The images correspond to air flow rates of $Q_{air}=0.018$, 0.083, and 1.0 $\frac{\mathrm{ml}}{\mathrm{s}}$, and time durations of 10 seconds for (d) and (e), and 3.3 seconds for (f). \label{fig:setup}}
\end{figure}

We studied the buoyancy driven flow generated in a small tank filled with fresh tap water. The rectangular prism shaped tank has its length 100 mm, its width is 50 mm, and it was filled with water up to a height of 100 mm (Fig.~\ref{fig:setup}a). Air bubbles were injected to the tank through a 1 mm diameter plastic tube at the bottom wall, located adjacent to one of the tank's side walls at its central cross section. The air was pumped into the tank at a precise volume flow rate by connecting the injection tube to a syringe pump fitted with either 20 or 60 ml syringes. The setup yielded discrete bubbles that rose in a narrow columnar region, confined to one of the sides of the tank. The bubbles in our experiment had an oblate spheroidal shape, with a horizontal to vertical aspect ration of approximately 2:1 with slight oscillations that they experienced as they were ascending in slightly undulating trajectories (Fig.~\ref{fig:setup}a and b).

The bubble injection at one side provided a strongly inhomogeneous buoyancy distribution, which generated an overturning, tank-wide circulating flow. As the air flow rate was changed, the bubbles appeared at different frequencies, and this allowed to control the buoyancy forcing. Indeed, the flow had complex features that whose characteristics depended on the air flow rate. At low flow rate the overturning flow was slow and the flow was laminar; importantly, and as will be discussed below, the flow is never really steady in this setup due to the discrete nature of the bubble forcing. At higher flow rate, the flow became unstable in the sense that the flow spatial structure became more convoluted and the intensity of the fluctuations grew rapidly with the forcing. This is demonstrated by the sequence of three flow visualization images shown in Fig.~\ref{fig:setup}d-f, and in supplementary movies 1 and 2.

\subsection{System parameters}\label{sec:params}

We define the characteristic bubble size as $D_{eq} \equiv \left(\volume_b/\frac{\pi}{6}\right)^{1/3} = 3.7\pm0.2$ mm, where $\volume_b$ is the bubble volume that was measured by stereo matching points along bubble perimeters through our 3D particle tracking system (Sec.~\ref{sec:PTV}). The bubble Prandtl number, defined as $\mathrm{Pr}_b = \nu / (L\,U_b)$, is equal to $3.8\times10^{-5}$ in this work, where $\nu$ is the water kinematic viscosity, and $U_b=260\,\pm$ mm/s is the bubble raise velocity that was measured through our particle tracking system. The bubble Reynolds number is thus $\mathrm{Re}_b = D_{eq}\,U_b / \nu = 950\pm50$, and the bubble Weber number is $\mathrm{We}_b = \rho U_b^2 \, D_{eq} / \sigma = 3.4 \pm 0.2$, where $\sigma$ is the water-air surface tension.

An important parameter in bubbly flows is the volume fraction of gas in the liquid. As bubbles were released locally in our system, we use an analogue of the volume fraction, defined as the mean inter-bubble vertical spacing divided by the bubble size. This parameter can be calculated as
\begin{equation}
\alpha = \frac{\text{bubble convective time}}{\text{bubble generation time}} = \frac{D_{eq} / U_b}{\volume_b / Q_{air}}
\label{eq:alpha}
\end{equation} 
and in this work $\alpha$ is in the range from 1\% to 54\%.

The flow in our system is driven by the buoyancy supplied through the air injection. As such, an analogy can be found with single phase buoyancy driven flows that are usually generated through temperature or solute non-uniformity, such as the Rayleigh-B\'{e}nnard convection.  
We therefore quantify the forcing in our system using an analogue Rayleigh number. In our bubble induced flow this can be defined as (cite climent and magnaudet 1999)
\begin{equation}
\mathrm{Ra} = \frac{g\,L^2\,\alpha}{\nu \, U_b}
\label{eq:Rayleigh}
\end{equation} 
and in our work it is in the range between $3.8\times 10^3$
and $2.1\times 10^5$. An increase in $\mathrm{Ra}$ can be interpreted as the forcing to the flow being intensifies. Furthermore, at low $\mathrm{Ra}$ the flow is laminar, while at higher $\mathrm{Ra}$ the flow destabilizes and the fluctuations intensify. The values of the various experimental parameters used in this work are outlined in Tab.~\ref{tab:params}.

\begin{table}[]
	\centering
	%\begin{tabularx}{0.55\textwidth}{c c c c c c}
	\begin{tabular}{>{\centering}p{2cm} >{\centering}p{2cm} >{\centering}p{2cm} >{\centering}p{2cm} >{\centering}p{2cm} >{\centering\arraybackslash}p{2cm}}
		\hline
		run & $Q_{air}$ [ml/s] & $\mathrm{Pr}_b$ & $\mathrm{Re}_b$ & $\alpha$ & $\mathrm{Ra}$ \\%  & $\mathrm{Ra}$ \\
		\hline
		1 & 0.018 & 3.8$\times 10^{-5}$ & 950 & 0.01 & 3.8$\times 10^{3}$ \\%& 187 \\
		2 & 0.028 & 3.8$\times 10^{-5}$ & 950 & 0.015 & 5.7$\times 10^{3}$ \\%& 226\\
		3 & 0.037 & 3.8$\times 10^{-5}$ & 950 & 0.02 & 7.6$\times 10^{3}$ \\%& 240\\
		4 & 0.055 & 3.8$\times 10^{-5}$ & 950 & 0.03 & 1.1$\times 10^{4}$ \\%& 242\\
		5 & 0.083 & 3.8$\times 10^{-5}$ & 950 & 0.045 & 1.7$\times 10^{4}$  \\%& 332\\
		6 & 0.11 & 3.8$\times 10^{-5}$ & 950 & 0.06 & 2.3$\times 10^{4}$ \\%& 378\\
		7 & 0.17 & 3.8$\times 10^{-5}$ & 950 & 0.09 & 3.4$\times 10^{4}$ \\%& 396\\
		8 & 0.22 & 3.8$\times 10^{-5}$ & 950  & 0.12 & 4.6$\times 10^{4}$ \\%& 591\\
		9 & 0.33 & 3.8$\times 10^{-5}$ & 950 & 0.18 & 6.8$\times 10^{4}$ \\%& 587\\
		10 & 0.50 & 3.8$\times 10^{-5}$ & 950 & 0.27 & 1.0$\times 10^{5}$ \\%& 826\\
		11 & 0.67 & 3.8$\times 10^{-5}$ & 950 & 0.36 & 1.4$\times 10^{5}$ \\%& 765\\
		12 & 1.00 & 3.8$\times 10^{-5}$ & 950 & 0.54 & 2.1$\times 10^{5}$ \\%& 908\\
		\hline
	\end{tabular}
	%\end{tabularx}
	\caption{Nominal system parameter values for the various experimental runs. The air flow rate column is given in milliliter of air injected by the syringe pump per second. \label{tab:params}}
\end{table}

\subsection{Data acquisition}

The flow was visualized and measured by adding Polyamide tracer particles, 55 $\micro$m in diameter (Dantec Dynamics) to the water at a moderate seeding density (approximately 1000 particles per frame were in the region of interest). The particles were illuminated from one side of the tank using a white LED pannel, and the light was reflected back into the tank through a mirror at the opposite end. Three digital cameras (Mikrotron, 1280$\times$1024), each fitted with a 60 mm lens, were placed in a configuration approximately perpendicular to the illumination direction, and looking towards the central region of the tankg (Fig.~\ref{fig:setup}c). The cameras were connected to an external DVR recording system that was also responsible for the synchronous camera triggering (IO industries). The cameras viewed the whole tank from one side to the other. Images were recorded are various rates starting from 30 Hz for the lowest air flow rate and up to 90 Hz for the highest flow rate. Data was recorded in separate runs, each lasting one minute. A total of 21 experimental runs were conducted, using 11 values of the air flow rate, starting from 0.018 ml/s and up to 1 ml/s. 
Flow visualizations using overlaid tracer particle images from this setup are shown in Fig.~\ref{fig:setup}d-f.

\subsection{3D particle tracking measurement}\label{sec:PTV}

The flow was measured by using the 3D particle tracking velocimetry method (3D-PTV). All of the experimental analysis was conducted through our in-house developed, open source software, MyPTV~\cite{Shnapp2022}. The region of interest in the measurements covered a volume in the center of the tank, of dimensions 70$\times$70$\times$40 mm, that allowed us to focus on the overturning circulation flow there. While the measurements used the images of the full tank, the particle tracking analysis was performed only on particles found in the region of interest described. In particular, the velocity measurements did not cover the region of the bubble column, but instead focused on the region of the overturning circulating flow (Fig.~\ref{fig:setup}a).

Before and after the experiment a calibration target, made of a single piece anodized Aluminum was placed in the tank. The target had 420 small calibration points distributed over three parallel planes. These dots were used to calibrate the cameras, by fitting the external and internal camera parameters, altogether 15 parameters. Our system, using our MyPTV open source software (cite), adopts the pinhole camera paradigm with a quadratic polynomial used to correct for non-linear aberrations. The particle positioning uncertainty in our system after this calibration scheme was approximately 0.15 mm, measured by the RMS of the calibration point deviations relative to their projections.

Following camera calibration we subtracted a static background from each experimental image, applied a Gaussian blur with standard deviation of 0.5 pixel, and segmented the tracer particles by thresholding and searching for bright spots. The segmented particles were then stereo matched to obtain 3D particle clouds using the ray traversal algorithm~\cite{Bourgoin2020}, and the particle clouds were tracked in time through the four-frame best estimate paradigm~\cite{Ouellette2005}. Finally, trajectories were smoothed my spline fitting with a window of 9 frames, that was also used to calculate the particle velocities and accelerations~\cite{Luethi2005}.

The experiments described yielded approximately $\sim\mathcal{O}(5\times 10^6)$ velocity samples for each flow rate, with $\sim\mathcal{O}(10^3)$ samples at each moment in time. There is approximately two minutes of data recorded per flow rate value (except for the highest flow rate for which 1 minute only is available) The spatial resolution, i.e. the typical spacing between simultaneous velocity sample, is approximately 5 mm.

\section{A buoyancy-wall friction equilibrium}~\label{sec:euilibrium_theory}

The flow in our system is driven by the buoyancy forces applied by the air bubbles. Further, assuming a steady state is achieved, or at least a quasi-steady state, this kinetic energy supply is balanced out by viscous dissipation of kinetic energy. Furthermore, if the flow in the tank is laminar, most of the viscous dissipation would occur in the boundary layer near the tank walls, because this is where the velocity gradients are the strongest. Thus, it is fair to approximate that at sufficiently low $\mathrm{Ra}$ the buoyancy energy input is balanced with the fluid friction in the boundary layer. This equilibrium assumption allows to derive a scaling law for the dependence of the Reynolds number of the overturning circulation flow.

As for the energy input, in a duration $T$ long enough for several bubbles to be released into the fluid, the work done on the fluid is equal to $W = F_B \, L \, T \, \frac{Q_{air}}{\volume_b}$, where $F_b$ is the average force acting on a single bubble; $T \, \frac{Q_{air}}{\volume_b}$ is the number of bubbles released during the time $T$ and $F_B \, L$ is the mechanical work done per bubble. The average force on the bubble can be calculated using Archimedes' principle as $F_B = \rho\,g\,\volume_b$ where $g$ is the gravitational acceleration and $\rho$ is the fluid density. Dividing the work done by $T$, and using eq.~\eqref{eq:alpha}, we get the mean energy injection rate by buoyancy into the flow as
\begin{equation}
E_{in} = \rho\,g\,\volume_b \cdot U_b \cdot \alpha \cdot \frac{L}{D_{eq}} \, .
\label{eq:Ein}
\end{equation}
The friction force along the tank walls can be estimated using boundary layer theory. For not too high Reynolds numbers it is fair to assume the boundary layer is laminar. This assumption leads to the well-known scaling of the wall shear stress, $\tau_w \equiv F_f / A = C \, \rho \,U^2 \,\mathrm{Re}^{-1/2}$~\cite{Schlichting2016}, where $F_f$ is the friction force, $A$ is the surface area, and
\begin{equation}
\mathrm{Re}=\frac{L\,U}{\nu}
\label{eq:Re}
\end{equation}
is the Reynolds number. In flows with known geometry, such as the flow above a flat plate, $U$ is the free stream velocity above the wall and $C$ is a constant that depends on the velocity profile shape. In the context of the flow in the tank studied here, the definition of $U$ is not straight forward, and the value of $C$ cannot be simply predicted; in fact, they might change slowly with $\mathrm{Ra}$ if the mean flow structure changes with $\mathrm{Ra}$. To proceed forward, we will define $U$ as the characteristic velocity scale associated with the overturning circulation flow in the bulk of the tank, giving an exact empirical definition later on (Sec.~\ref{sec:results_circulation}). Furthermore, for the sake of comparison, let us recall that in the Blasius boundary layer profile, $C \approx 1.32$~\cite{Schlichting2016}. Overall, the rate of energy loss due to the friction of the fluid against the tank walls is   
\begin{equation}
E_{out} = C \, \rho \,U^3 \,\mathrm{Re}^{-1/2} \, A_{tank} \, .
\label{eq:Eout}
\end{equation}
where $A_{tank}$ is the wetted surface area of the tank walls.

Following through with the equilibrium assumption, by equating eq.~\eqref{eq:Ein} and \eqref{eq:Eout} and rearranging the terms we obtain that 
\begin{equation}
\mathrm{Re}^{5/2} = \frac{g\,U_b\,\volume_b\,\alpha\,L^4}{D_{eq} \, \nu^3 \,A_{tank} \, C} \, .
\end{equation}
We can also express the bubble volume as $\frac{\pi}{6}\,D_{eq}$ according to the definition in Sec.~\ref{sec:params}, and in our case $A_{tank} = \frac{7}{2}\,L^2$ ($L$ being the length of our tank, Sec.~\ref{sec:methods}). Rearranging this expression by using eq.~\eqref{eq:Rayleigh} and the definition of the bubble Reynolds number, $\mathrm{Re}_b$, we obtain the following relation
\begin{equation}
\mathrm{Re} = A \, \mathrm{Ra}^{2/5} \, \mathrm{Re}^{4/5}_b \, .
\label{eq:equilibrium_scaling}
\end{equation}
In the experimental tank used in this study, $A = (\frac{\pi}{21\, C})^{2/5}$. As long as the structure of the mean circulation flow will not vary with $\mathrm{Ra}$ and as long as the laminar boundary layer assumption holds the value of $C$ will be constant, and thus $A$ will not change with $\mathrm{Ra}$. Due to the ambiguity in the definition of $U$, the numerical value of $A$ cannot be determined apriori without knowing the full flow field.

Scaling laws in which $\mathrm{Re} \sim \mathrm{Ra}^{2/5}$ are not uncommon in buoyancy driven flows. For example, the same scaling exponent is found also in low $\mathrm{Pr}$, medium $\mathrm{Ra}$ numbers in the Rayleigh-B\'{e}nnard convection (Regime II in Ref.~\cite{Grossmann2000}), and in horizontal convection, in which an overturning circulation is driven by a horizontal temperature gradient~\cite{Hughes2008}. Indeed, the $\frac{2}{5}$ exponent will pop up in every case in which the forcing is independent of the velocity while the dissipation occurs due to friction in a laminar boundary layer. In our case, the $U$ independence of the forcing results from using the bubble rise velocity in the calculation of $E_{in}$ in eq.~\eqref{eq:Ein}. In other words, we assumed that the bubble velocity does not change with $U$ nor with $\mathrm{Ra}$, which is reasonable since the bubble rise velocity is more than an order of magnitude larger than the range of $U$ investigated here (see Sec.~\ref{sec:results}).

% ==============================================================
%
%          Results
%
% ==============================================================

\section{Results} \label{sec:results}

\subsection{Tank-scale circulation} \label{sec:results_circulation}

To test eq.~\eqref{eq:equilibrium_scaling}, we first considered the characteristic velocity scale of the mean overturning circulation in the tank. For that, we used the volumetric average of the temporal mean flow field
\begin{equation}
U \equiv \frac{1}{\volume_{\mathrm{m.v.}}}\iiint \limits_{\mathrm{m.v.}} |\mathbf{\avt{u}}| \, dV \, .
\label{eq:U}
\end{equation} 
where $\avt{\mathbf{u}}(\vec{x})$ is the time averaged flow velocity field and $\volume_\mathrm{m.v.}$ is whole the measurement volume. This particular definition of $U$ was motivated by the relation it has to the mean flow kinetic energy per unit mass, $E=\frac{1}{2}U^2$, as the argument leading to eq.~\eqref{eq:equilibrium_scaling} is essentially an energy balance.    
In practice, calculating $\avt{\mathbf{u}}(\vec{x})$ from our Lagrangian particle tracking measurement requires some manipulation as the velocity samples in this method are taken at the positions the particle happen to be. Thus, the mean Eulerian velocity field was calculated by binning the particle velocities according to their positions and averaging the binned samples. For the binning, we used a grid of 900 voxels with volumes of 100 mm$^3$ each. Following that, $U$ was calculated with eq.~\eqref{eq:U} by numerical integration. Then, after calculating the characteristic velocity the flow Reynolds number is calculated using eq.~\eqref{eq:Re}.

\begin{figure}[t]
	\centering
	\includegraphics[width=\textwidth]{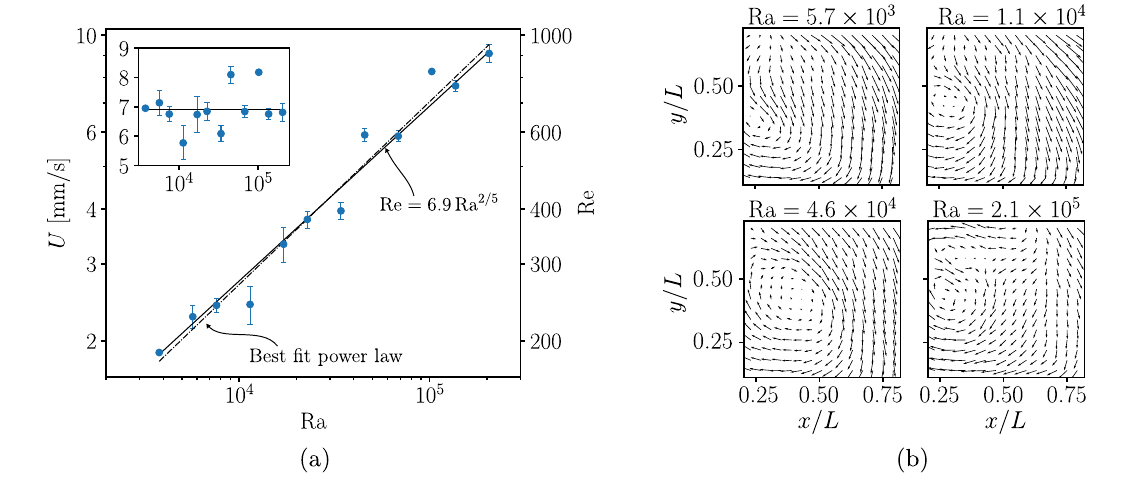}
	\caption{(a) The characteristic velocity, defined according to eq.~\eqref{eq:U}, and the corresponding Reynolds number, are plotted against the Rayleigh number in log-log scales. A fit to the data with respect to eq.~\eqref{eq:equilibrium_scaling} is shown as a straight line and results are shown on the graph. A best fit power law to the data with $\mathrm{Ra} \propto \,\mathrm{Re}^a$ is also shown where $a=0.41$ was obtained with a least squares minimization. The inset shows the same data, divided by $\mathrm{Ra}^{2/5}$, thus showing the scatter in the estimation of the coefficient $A=6.9$ in eq.~\eqref{eq:equilibrium_scaling}. Error bars were calculated by dividing the data into two time-based sub-samples (first and last half) and repeating the calculation on each. (b) Quiver plots showing two dimensional cross sections of the mean velocity field at four Rayleigh number values. \label{fig:mean_flow}}
\end{figure}

In Fig.~\ref{fig:mean_flow}a, $U$ and $\mathrm{Re}$ are plotted against the Rayleigh number in log-log scales. The error bars in the figure represent the statistical convergence of the calculation of the time averages, as estimated by dividing the dataset into two time-based sub-samples and repeating the calculation over each. Using least squares fitting to the data with respect to eq.~\eqref{eq:equilibrium_scaling}, the $\mathrm{Re}^{1/2} = 6.9\,\mathrm{Ra}^{2/5}$ power law, shown as a continuous line in the figure, was obtained. The fit is seen to capture the trend well, although the scatter around the fit, which is highlighted also in the inset, is larger that the error bars.

As the error bars in the figure that correspond to statistical convergence are smaller than the scatter, a physical explanation is needed. As discussed in Sec.~\ref{eq:equilibrium_scaling}, the coefficient $A$ in eq.~\eqref{eq:equilibrium_scaling} is sensitive to the structure of the mean flow - changes in the patterns of the streamlines in different experimental repetitions would effectively change the value of $A$. Such deviations in the flow structure could result from undulations of the bubbles' raising trajectories. This could affect the mean flow pattern, which in a feed back loop with the bubbles' trajectory undulations will further increase the deviations in the mean flow from one repetition to the other. The mean flow fields illustrated in four different experimental runs in Fig.~\ref{fig:mean_flow}b demonstrate this issue. Indeed, although in all cases shown an overturning circulating patterns could be identified, the details of the flow, such as the location of the center of rotation seen in each case, differ somewhat from one case to another. These changes mean that the mean flow fields from each repetition are not self-similar, and they can affect value of $A$ in each run. Effectively, this mechanism means that ergodicity in this flow is not ensured, meaning that the convergence to the time average does not imply the true mean value is resolved. This is consistent with the scatter in Fig.~\ref{fig:mean_flow} being larger than the error bars.

Despite the scatter observed, a trend in the data could be observed over the two orders of magnitude in $\mathrm{Ra}$ range used. Thus, to reaffirm that the $\frac{2}{5}$ power law was not imposed by the previous fit, another fit to the data was performed while keeping the exponent of the Rayleigh number free, namely $\mathrm{Re}^{1/2} = a\,\mathrm{Ra}^{b}$. The results, shown as the "best fit power law" in Fig.~\ref{fig:mean_flow}, gave an exponent of $b=0.41$. The value is very close to the theoretical prediction of $\frac{2}{5}$. Overall, this result suggests that the equilibrium argument and eq.~\eqref{eq:equilibrium_scaling} are indeed consistent with our results.

From the coefficient multiplying  the $\mathrm{Ra}^{2/5}$ fit, the average value of $A$ could be estimated. Rearranging eq.~\eqref{eq:equilibrium_scaling} and using the results of the data fit we obtain $A = (6.9\pm0.65)\,\mathrm{Re}_b^{-4/5} = 0.029 \pm 0.003$, where the uncertainty was taken as the standard deviation of the residuals shown in the inset of Fig.~\ref{fig:mean_flow}. In addition to that the average value of the boundary layer coefficient, $C=(9.5\pm2.5)\times10^{-4}$, can also be calculated. It is emphasized here, again, however, that the values of these coefficients are expected to depend on the geometry of the experimental setup used.

\subsection{The growth of fluctuations and emergence of unstable flow}

With the intensification of the bubble injection the Rayleigh number and the Reynolds number grow. While the flow at the low values of bubble injection was laminar, as $\mathrm{Ra}$ was increased fluctuations began to appear and grow and the structure of the (instantaneous) flow became more complex. In this section we will characterize the fluctuating component of the flow.

\begin{figure}[h!]
	\centering
	\includegraphics[width=\textwidth]{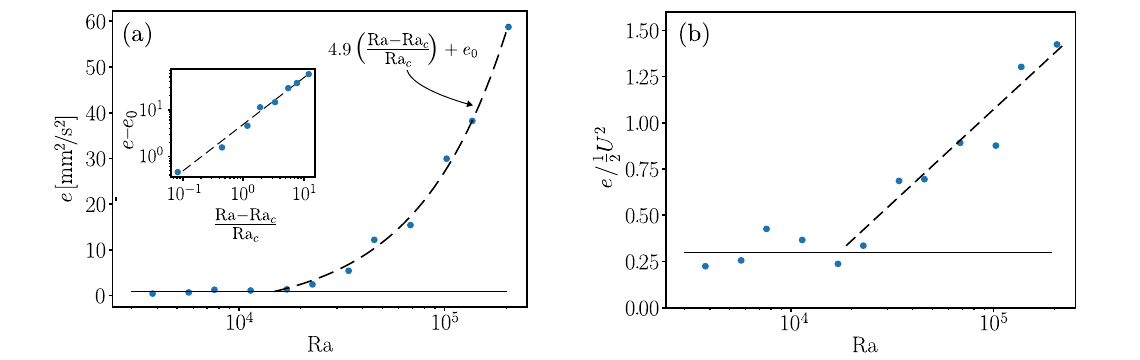}
	\caption{(a) The kinetic energy of the fluctuations is plotted as a function of the Rayleigh number in log-log scales. A continuous line shows a base level of the fluctuation energy, $e_0=0.84$ mm$^2$/s$^2$. A linear growth of the fluctuations above a critical Rayleigh number, $\mathrm{Ra}_c = (1.6\pm0.1)\times 10^{4}$ is shown as a dashed line. (b) The intensity of the fluctuations relative to the mean flow energy is shown as a function of the Rayleigh number. The baseline intensity of 0.3 is marked by the continuous horizontal line. \label{fig:transition}}
\end{figure}

The strength of the fluctuations is characterized by considering the mean kinetic energy per unit mass of the fluctuating component of the flow. Specifically, the variance of the $i$'th velocity component at each point is defined as $\sigma_i^2 = \avt{\left(u_i - \avt{u_i}\right)^2}$, and this allows to define the fluctuation strength as
\begin{equation}
e \equiv \frac{1}{\volume_{\mathrm{m.v.}}}\iiint \limits_{\mathrm{m.v.}} \frac{1}{2} (\sigma_x^2 + \sigma_y^2 + \sigma_z^2) \, dV \, .
\end{equation}
In practice, the variances, $\sigma_i$, were calculated by binning the velocity samples in space and calculating the variance of the samples in each bin, followed by the numerical volume integration.

The fluctuation strength is shown as a function of the Rayleigh number in Fig.~\ref{fig:transition}a. Even at the smallest Rayleigh number used, for which $\mathrm{Re}\approx 190$ and the flow was laminar, the fluctuation energy was not zero. Temporal fluctuations in this flow are expected even at very low $\mathrm{Re}$ due to the  discrete nature of energy injection in this setup. Indeed, bubbles in this system are released in discrete events, where at the lowest air flow rates single bubbles rose through the tank. This can be understood by noting the low values of $\alpha$, that reached as low as 1\% (Tab.~\ref{tab:params}). Thus, the forcing applied on the flow for all $\mathrm{Ra}$ was inherently time dependent. Therefore, there is no wonder that even at the laminar flow regime time variations of the flow existed. 

In the low $\mathrm{Ra}$ (defined empirically here as $\mathrm{Ra}\leq1.1\times10^4$) range of our measurements no significant change of the fluctuation magnitude with $\mathrm{Ra}$ was detected. Thus, the level of the fluctuation strength at the low $\mathrm{Ra}$ regime, $e_0=0.84$ mm$^2$/s$^2$, was calculated as the average of $e$ in the first four experimental runs. This value of $e_0$ is approximately 30\% of the mean flow energy at this Rayleigh number range, as seen in Fig.~\ref{fig:transition}b.

While in the low $\mathrm{Ra}$ range of our measurements the fluctuation kinetic energy was approximately constant, at higher $\mathrm{Ra}$ the fluctuation energy increased rapidly with $\mathrm{Ra}$ (Fig.~\ref{fig:transition}). This corresponds to a transition from the regime of constant $e \approx e_0$ to a regime in which $e$ is growing with the Rayleigh number significantly. As seen in Fig.~\ref{fig:transition}b, the fluctuation kinetic energy in the second regime reached up to values of approximately 150\% of the kinetic energy of the mean flow ($\frac{1}{2}U^2$) at the highest $\mathrm{Ra}$ case. Indeed, at the highest $\mathrm{Ra}$ the fluctuations were significantly stronger than the mean overturning flow.

To characterize the transition between the two regimes we fit an empirical power law
\begin{equation}
e = 
\begin{cases}
e_0 \qquad &\text{for } \mathrm{Ra}<\mathrm{Ra}_c \\
\beta \, (\mathrm{Ra} - \mathrm{Ra}_c)^\gamma \qquad &\text{for } \mathrm{Ra}_c\leq \mathrm{Ra} 
\end{cases}
\end{equation}
to the data using least squares fitting, where $\mathrm{Ra}_c$ is a critical Rayleigh number for which the transition occurred. The results of the fit gave $\mathrm{Ra}_c = (1.6\pm0.1)\times10^4$ for the transition value. Furthermore, the fit gave an exponent of $\gamma=1.05\pm0.05$, which corresponds to slightly faster than linear growth of the fluctuation energy above the critical Rayleigh number. The data and a linear fit above the transition are shown in log-log scales against the order parameter $\frac{\mathrm{Ra}-\mathrm{Ra}_c}{\mathrm{Ra}_c}$ in the inset of Fig.~\ref{fig:transition}a, showing good agreement with the data over two orders of magnitude in $\mathrm{Ra}_c$ above the transition.

\begin{figure}[h!]
	\centering
	\includegraphics[width=\textwidth]{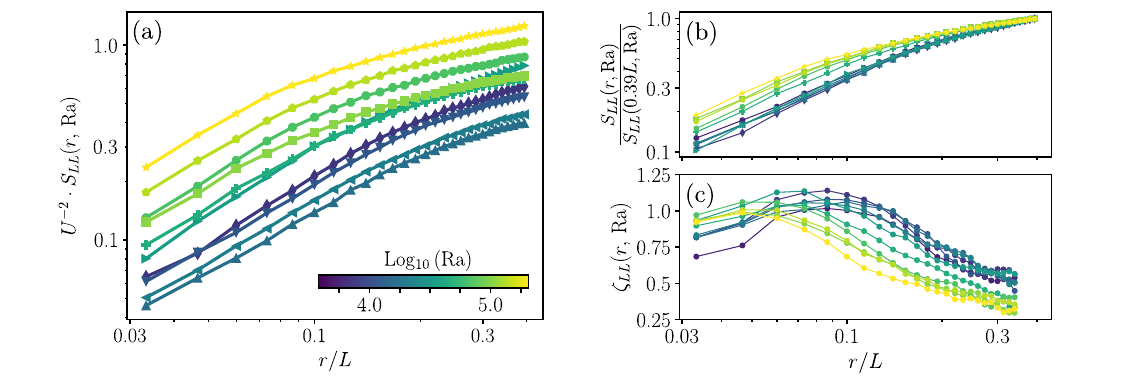}
	\caption{(a) The second order longitudinal structure function, normalized using the characteristic velocity, is shown as a function of scale normalized by the tank length for various Rayleigh number values. (b) The second order longitudinal structure function, divided by the values at the respective Rayleigh number and the largest separation distance, $r=0.39L$. (c) The local (logarithmic) slope of the second order longitudinal structure function shown as a function of scale normalized by the tank length. \label{fig:SLL}}
\end{figure}

The transition of the flow to a different state was observed also in the topology of the flow. A qualitative observation of this transition can be observed in the flow visualization images in Fig.~\ref{fig:setup}d, e and f. Below the transition the flow topology is smooth where the particle streaks clearly outline the large scale circulation, while above the transition many of the particles exhibit helical shapes that could indicate the existence of vortical structures at scales much smaller than $L$. This qualitative view of the flow topology is also seen in supplementary movies 1 and 2 at $\mathrm{Ra}=7.6\times10^3$ and $\mathrm{Ra}=2.1\times10^5$ respectively.

To characterize the change in the flow structure more quantitatively we consider the longitudinal velocity spatial difference, defined as
\begin{equation}
	\delta u_r(\vec{x},\, r,\, t,\,\mathrm{Ra}) = \Big[ u(\vec{x},\, t, \,\mathrm{Ra}) - u(\vec{x} +\vec{r},\, t,\,\mathrm{Ra}) \Big] \cdot \frac{\vec{r}}{r} \, \, ,
\end{equation}
where we examine its second moment,  
\begin{equation}
S_{LL}(r, \, \mathrm{Ra}) = \av{ \, \avt{ \rule{0em}{.9em} \delta u_r(\vec{x},\, t,\,\mathrm{Ra})^2} \, } \, .
\end{equation}
Here, $S_{LL}$ is defined by averaging over all samples, as implied by the time ($\avt{\,\cdot\,}$) and space ($\av{\cdot}$) averages. In turbulence, $S_{LL}$ is analogous to the second order Eulerian structure function~\cite{Pope2000}. The difference between $S_{LL}$ and the structure functions is that in inhomogeneous flows, such as the one considered here, the structure function depends on space, namely, it is a function of $\vec{x}$. However, here we average over $\vec{x}$ in order to obtain a macroscopic observable, as this will suffice to demonstrate our point. Thus, $S_{LL}$ in this work is a function of the separation distance and the Rayleigh number only. Being the second moment of the spatial relative velocities, higher values of $S_{LL}$ at some point as compared to another can be interpreted as having more kinetic energy associated with the scale $r$ at Rayleigh number $\mathrm{Ra}$.

In Fig.~\ref{fig:SLL}a $S_{LL}(r, \,\mathrm{Ra})$ is shown for the various experimental runs, where it is normalized by $U^2$. For all $\mathrm{Ra}$ cases the curves are seen to increase with $r$, indicating that more kinetic energy is associated with the large scales of the flow as compared to smaller ones. Furthermore, similar to Fig.~\ref{fig:transition}b, a trend is seen for the normalized $S_{LL}$ curves increase with $\mathrm{Ra}$; scatter around this trend exists, presumably due to the same mechanism that led to the scatter of $A$ in Fig.\ref{fig:mean_flow}a. This demonstrates that the growth of the kinetic energy with the forcing ($\mathrm{Ra}$) is not confined to a single scale, but instead it occurs at all the scales available for our measurement.

To focus in on the appearance of small flow scales with the growth of $\mathrm{Ra}$, we show $S_{LL}$ for various $\mathrm{Ra}$, normalized by its values at the largest scale available for our measurement in Fig.~\ref{fig:SLL}b. As $\mathrm{Ra}$ grows, the kinetic energy of the small scales relative the kinetic energy of large scales increases. Specifically, in the range of our measurement this fraction increases approximately from 0.1 to 0.2 at the smallest scale available, namely an increase by a factor of 2. The same observation can also be seen in Fig.~\ref{fig:SLL}c, which shows the local logarithmic slope of $S_{LL}$
\begin{equation}
\zeta_{LL}(r,\,\mathrm{Ra}) = \frac{\partial \, (\log \, S_{LL})}{\partial \, (\log \, r)} \,\, .
\end{equation}
Positive values of $\zeta_{LL} $ indicate that the kinetic energy grows with the scale, $r$, while negative values indicate it decreases with the scale. Also, higher positive values indicate that the energy grows faster with $r$ as compared to a lower value. 
For all $\mathrm{Ra}$ the local slope is approximately 1 at small $r$, and it decreases as $r$ increases, meaning that the kinetic energy at larger scales grows slower with $r$. Furthermore, for the larger $\mathrm{Ra}$ cases the slopes decrease faster and their decrease begins at smaller $r$ values. This indicates that for the higher $\mathrm{Ra}$, i.e. above the transition, the kinetic energy is spread more evenly across the scales, namely that a larger fraction of total the kinetic energy in the flow is present at smaller scales as compared to the lower $\mathrm{Ra}$ case. Overall, these observations demonstrate that as the Rayleigh number grew above the transition,  more and more kinetic energy began to accumulate in the smaller scales of the flow. Notably, in the inertial range of homogeneous isotropic turbulence, Kolmogorov theory predicts $\zeta_{LL} = 2/3$~\cite{Pope2000}; the lack of this plateau (or any other) in our measurements indicate the clear absence of an inertial range in our experiment, as expected in such a low Reynolds number flow.

\section{Discussion \& conclusions} \label{sec:conclusinos}

This work presents an experimental characterization of the flow driven by buoyant bubbles rising at moderate Reynolds numbers in a water tank. Despite the seemingly simple settings, the flow presents complex dynamics. The first feature is the appearance of a tank-scale circulation flow. The circulation is balanced by the wall friction, which leads to a power law of $\mathrm{Re}\sim\mathrm{Ra}^{2/5}$ (eq.~\eqref{eq:equilibrium_scaling}), as confirmed in our measurements (Fig.~\ref{fig:mean_flow}). Due to the discrete nature of the forcing the circulation is unsteady and exhibits slow temporal fluctuations with an intensity of approximately 30\% of the mean circulation. Furthermore, the topology of the mean circulation was seen, through flow visualizations, to be rather complex and three dimensional, although this was not explored in depth. As the Rayleigh number was increased (by increasing the bubble injection rate), a transition to a state of fluctuation growth occurred. Above the this transition, the energy of the fluctuating flow component grew  with the Rayleigh number scaling of $\mathrm{Ra}^\gamma$ with $\gamma=1.05\pm0.05$, reaching approximately 150\% of the mean circulating flow energy at the highest Rayleigh number tested. Increasing Rayleigh number above the transition caused increasing fractions of the fluctuation energy to appear in smaller scales; this was seen in the emergence of helical trajectories in the flow visualizations (Fig.~\ref{fig:mean_flow}e-f, and supplementary movies 1 and 2). Notably, despite the observed instability and growth of fluctuations, turbulence was never fully developed in the sense that no inertial range is found in the structure functions (Fig.~\ref{fig:SLL}). This is in concurrence with the low relatively low Reynolds numbers in our experiment.

One of the central questions left open is what is the mechanism responsible for the observed transition? Several possible explanations come to mind. One possible explanation is that agitation in the bubble wake~\cite{Riboux2013, Rensen2005} is entrained into the bulk of the tank, however, from Fig.~\ref{fig:SLL}b it is seen that the increased fluctuation energy begins at scales larger than the bubble diameter ($>0.1\frac{r}{L}$), making this mechanism less probable. Another possible explanation is that the fluctuations occur due to an instability of the large scale circulation when $\mathrm{Re}$ is sufficiently high. However, the Reynolds number at the transition in our experiment was approximately $\mathrm{Re} \approx 320$ (at $\mathrm{Ra}_c$ according to eq.~\eqref{eq:equilibrium_scaling}), which seems too low for this mechanism to be active. A mechanism for the instability that does seems to align with our system configuration is that the bubbles rising at the side of the tank generate a jet of rising liquid above the tube from which the bubbles are released, and that then, vorticity is generated and fluctuations are amplified by the jet impingement on the free surface. In this region, the upwards directed liquid jet is forced to make a sharp turn thus producing angular momentum and vorticity. Furthermore, after the first turn, the flow impinges on the right-side tank wall, producing Taylor-G\"{o}rtler-like vortical structures. This mechanism is consistent with the flow visualizations in Fig.~\ref{fig:setup}d-f and supplementary movies 1 and 2, where spiraling particle tracks are seen to eject from the region in which the bubble column reaches the free surface, and helical trajectories can be clearly seen across the wall opposite to the bubble column. This scenario is analogous to the instability that occurs in open cavity flows, in which the shear layer at the open end of the cavity impinges on the downstream cavity wall, causing amplification of disturbances in the shear layer~\cite{Rockwell1978, Gharib1987} and Taylor-G\"{o}rtler like structures on the opposite wall~\cite{Koseff1984}.

The nature of this system, although quite simple from a configuration point of view, is able to produce quite a complex physical behavior. First, the forcing applied on the flow is "free" in the sense that the flow in the tank has a feedback effect on the bubble column. This could change the system behavior in complex, unexpected ways, as for example was observed by the scatter of the $A$ value in Fig.~\ref{fig:mean_flow}. An analogous kind of feedback loop occurs in cavity flows, where interactions of the free shear layer with the flow impingement on the downstream cavity wall is known to produce coherent acoustic oscillations through a feed back mechanism~\cite{Rockwell1978}. Second, the forcing in our bubble system is applied in a temporally discrete fashion, so oscillations and fluctuations can occur already at very low Reynolds numbers. Furthermore, the fluid agitation in the bubble wakes, that is known to lead to pseudo-turbulence states in bubbly flows~\cite{Rensen2005, Riboux2013, Almeras2015, Risso2018, Innocenti2021} could make the problem even more difficult, as energy is injected to the system at various scales. All in all, these issues are expected to lead to significant difficulties should low-order modeling of this system be attempted.

The system under investigation in this study presents several interesting features, including flow oscillations and fluctuations at low Reynolds numbers, and a transition to a regime of highly disturbed flow. The potential to manipulate these features could be utilized in applications such as bio-reactors used for growing aquatic organisms, such as algae and plankton. In these systems, bubbles play a crucial role in oxygenation and nutrient mixing. By precisely adjusting air flow rates, one could potentially achieve desired mixing rates while ensuring that the flow remains below the transition point. This strategy could be leveraged to prevent excessive flow fluctuations that could potentially harm the organisms being cultivated in the reactor. For the flow conditions tested here, velocity fluctuations are moderate and representative of turbulence levels in the ocean~\cite{Franks2021}. This setup could thus also serve as a platform to systematically investigate interactions between turbulence and phyto- or zooplankton.

There are numerous parameters that could influence the behavior of the system in addition to the air flow rate that was used here: from the bubble size to the aspect ratio of the tank and the position of the bubble injector. Indeed, studying the phase space of this system holds great potential for other flow regimes to be discovered. This question, and several others, are thus left for future investigations.

\section*{Acknowledgments}

We are grateful to Marius Neamtu Halic, Stefano Brizzolara, and Alessandro Gambino for fruitful discussions. RS was a Rothschild postdoctoral fellow during the initial parts of this work.

\section*{Declaration of Interests}

The authors report no conflict of interest.

%\appendix

\bibliography{bibliography}

\begin{thebibliography}{10}

\bibitem{Thorpe1992}
S.~A. Thorpe.
\newblock Bubble clouds and the dynamics of the upper ocean.
\newblock {\em Quarterly Journal of the Royal Meteorological Society},
  118(503):1--22, January 1992.

\bibitem{Harteveld2003}
W.~K. Harteveld, R.~F. Mudde, and H.~E. A. Van~Den Akker.
\newblock Dynamics of a bubble column: Influence of gas distribution on
  coherent structures.
\newblock {\em The Canadian Journal of Chemical Engineering}, 81(3-4):389--394,
  2003.

\bibitem{Oresta2009}
P.~Oresta, R.~Verzicco, D.~Lohse, and A.~Prosperetti.
\newblock Heat transfer mechanisms in bubbly rayleigh-b{\'{e}}nard convection.
\newblock {\em Physical Review E}, 80(2):026304, 2009.

\bibitem{Mezui2022}
Y.~Mezui, M.~Obligado, and A.~Cartellier.
\newblock Buoyancy-driven bubbly flows: scaling of velocities in bubble columns
  operated in the heterogeneous regime.
\newblock {\em Journal of Fluid Mechanics}, 952, 2022.

\bibitem{Liu2010}
N.~Liu, Q.~Zhang, G.-L. Chin, E.-H. Ong, J.~Lou, C.-W. Kang, W.~Liu, and
  E.~Jordan.
\newblock Experimental investigation of hydrodynamic behavior in a real
  membrane bio-reactor unit.
\newblock {\em Journal of Membrane Science}, 353(1–2):122--134, May 2010.

\bibitem{Lohse2010}
D.~Lohse and K.-Q. Xia.
\newblock Small-scale properties of turbulent {R}ayleigh-{B}\'{e}nard
  convection.
\newblock {\em Annual Review of Fluid Mechanics}, 42(1):335--364, 2010.

\bibitem{Hughes2008}
Graham~O. Hughes and Ross~W. Griffiths.
\newblock Horizontal convection.
\newblock {\em Annual Review of Fluid Mechanics}, 40(1):185--208, 2008.

\bibitem{Lohse2018}
D.~Lohse.
\newblock Bubble puzzles: From fundamentals to applications.
\newblock {\em Phys. Rev. Fluids}, 3:110504, 2018.

\bibitem{Almeras2015}
E.~Alm{\'{e}}ras, F.~Risso, V.~Roig, S.~Cazin, C.~Plais, and F.~Augier.
\newblock Mixing by bubble-induced turbulence.
\newblock {\em Journal of Fluid Mechanics}, 776:458--474, 2015.

\bibitem{Mathai2020}
V.~Mathai, D.~Lohse, and C.~Sun.
\newblock Bubbly and buoyant particle{\textendash}laden turbulent flows.
\newblock {\em Annual Review of Condensed Matter Physics}, 11(1):529--559,
  2020.

\bibitem{Risso2018}
F.~Risso.
\newblock Agitation, mixing, and transfers induced by bubbles.
\newblock {\em Annual Review of Fluid Mechanics}, 50(1):25--48, 2018.

\bibitem{Innocenti2021}
A.~Innocenti, A.~Jaccod, S.~Popinet, and S.~Chibbaro.
\newblock Direct numerical simulation of bubble-induced turbulence.
\newblock {\em Journal of Fluid Mechanics}, 918, 2021.

\bibitem{Gong2009}
X.~Gong, S.~Takagi, and Y.~Matsumoto.
\newblock The effect of bubble-induced liquid flow on mass transfer in bubble
  plumes.
\newblock {\em International Journal of Multiphase Flow}, 35(2):155--162, 2009.

\bibitem{Mezui2023}
Y.~Mezui, M.~Obligado, and A.~Cartellier.
\newblock Buoyancy-driven bubbly flows: role of meso-scale structures on the
  relative motion between phases in bubble columns operated in the
  heterogeneous regime.
\newblock {\em Journal of Fluid Mechanics}, 962, may 2023.

\bibitem{Caballina2003}
O.~Caballina, E.~Climent, and J.~Du{\v{s}}ek.
\newblock Two-way coupling simulations of instabilities in a plane bubble
  plume.
\newblock {\em Physics of Fluids}, 15(6):1535--1544, 2003.

\bibitem{Simiano2006}
M.~Simiano, R.~Zboray, F.~de~Cachard, D.~Lakehal, and G.~Yadigaroglu.
\newblock Comprehensive experimental investigation of the hydrodynamics of
  large-scale, {3D}, oscillating bubble plumes.
\newblock {\em International Journal of Multiphase Flow}, 32(10-11):1160--1181,
  2006.

\bibitem{Climent1999}
E.~Climent and J.~Magnaudet.
\newblock Large-scale simulations of bubble-induced convection in a liquid
  layer.
\newblock {\em Physical Review Letters}, 82(24):4827--4830, 1999.

\bibitem{Ruzicka2003}
M.C Ruzicka and N.H Thomas.
\newblock Buoyancy-driven instability of bubbly layers: analogy with thermal
  convection.
\newblock {\em International Journal of Multiphase Flow}, 29(2):249--270, feb
  2003.

\bibitem{Nakamura2021}
K.~Nakamura, H.~N. Yoshikawa, Y.~Tasaka, and Y.~Murai.
\newblock Bifurcation analysis of bubble-induced convection in a horizontal
  liquid layer: role of forces on bubbles.
\newblock {\em Journal of Fluid Mechanics}, 923, 2021.

\bibitem{Rockwell1978}
D.~Rockwell and E.~Naudascher.
\newblock Review - {Self}-sustaining oscillations of flow past cavities.
\newblock {\em Journal of Fluids Engineering}, 100(2):152--165, 1978.

\bibitem{Koseff1984}
J.~R. Koseff and R.~L. Street.
\newblock Visualization studies of a shear driven three-dimensional
  recirculating flow.
\newblock {\em Journal of Fluids Engineering}, 106(1):21--27, 1984.

\bibitem{Shnapp2019}
R.~Shnapp, E.~Shapira, D.~Peri, Y.~Bohbot-Raviv, E.~Fattal, and A.~Liberzon.
\newblock Extended 3d-{PTV} for direct measurements of lagrangian statistics of
  canopy turbulence in a wind tunnel.
\newblock {\em Scientific Reports}, 9(1), 2019.

\bibitem{Shnapp2022}
R.~Shnapp.
\newblock {MyPTV}: A python package for 3d particle tracking.
\newblock {\em Journal of Open Source Software}, 7(75):4398, 2022.

\bibitem{Bourgoin2020}
Mickaël Bourgoin and Sander~G. Huisman.
\newblock Using ray-traversal for {3D} particle matching in the context of
  particle tracking velocimetry in fluid mechanics.
\newblock {\em Review of Scientific Instruments}, 91(8), August 2020.

\bibitem{Ouellette2005}
Nicholas~T. Ouellette, Haitao Xu, and Eberhard Bodenschatz.
\newblock A quantitative study of three-dimensional {Lagrangian} particle
  tracking algorithms.
\newblock {\em Experiments in Fluids}, 40(2):301--313, 2005.

\bibitem{Luethi2005}
B.~L\"{u}thi, A.~Tsinober, and W.~Kinzelbach.
\newblock Lagrangian measurement of vorticity dynamics in turbulent flow.
\newblock {\em Journal of Fluid Mechanics}, 528:87--118, 2005.

\bibitem{Schlichting2016}
H.~Schlichting and K.~Gersten.
\newblock {\em Boundary-layer theory}.
\newblock springer, 2016.

\bibitem{Grossmann2000}
S.~Grossmann and D.~Lohse.
\newblock Scaling in thermal convection: a unifying theory.
\newblock {\em Journal of Fluid Mechanics}, 407:27--56, 2000.

\bibitem{Pope2000}
S.~B. Pope.
\newblock {\em Turbulent Flows}.
\newblock Cambridge University Press, 2000.

\bibitem{Riboux2013}
G.~Riboux, D.~Legendre, and F.~Risso.
\newblock A model of bubble-induced turbulence based on large-scale wake
  interactions.
\newblock {\em Journal of Fluid Mechanics}, 719:362--387, 2013.

\bibitem{Rensen2005}
J.~Rensen, S.~Luther, and D.~Lohse.
\newblock The effect of bubbles on developed turbulence.
\newblock {\em Journal of Fluid Mechanics}, 538(-1):153, 2005.

\bibitem{Gharib1987}
M.~Gharib and A.~Roshko.
\newblock The effect of flow oscillations on cavity drag.
\newblock {\em Journal of Fluid Mechanics}, 177:501--530, 1987.

\bibitem{Franks2021}
P.~J.~S. Franks, B.~G. Inman, J.~A. MacKinnon, M.~H. Alford, and A.~F.
  Waterhouse.
\newblock Oceanic turbulence from a planktonic perspective.
\newblock {\em Limnology and Oceanography}, 67(2):348--363, 2021.

\end{thebibliography}
\bibliographystyle{unsrt}

\end{document}